\documentclass[12pt,twocolumn]{spieman}  
\usepackage{amsmath,amsfonts,amssymb}
\usepackage{graphicx}
\usepackage{setspace}
\usepackage{tocloft}
\usepackage{lineno}
\usepackage{hypernat}
\usepackage[nolist]{acronym}

\usepackage{./styles/shortcuts} 
\usepackage{import}      
\usepackage{svg}
\usepackage{fancyhdr}

\begin{acronym}
  \acro{SM}{single mode}
  \acro{MM}{multi-mode}
  \acro{GRIN}{gradient index}
\end{acronym}

\title{Characterization of Gradient Index Fibers}

\author[a]{Robert Leonard}
\author[b,c]{Matthew Marshall}
\author[d]{Seth Hyra}
\author[c,e]{Meagan Plummer}
\author[c]{Jacob Williamson}
\author[d,*]{Spencer Olson}

\affil[a]{Space Dynamics Laboratory, Quantum Sensing \& Timing, North Logan, UT 84341, USA}
\affil[b]{Department of Physics and Astronomy, University of New Mexico, Albuquerque, NM 87131, USA}
\affil[c]{Universities Space Research Association}
\affil[d]{Air Force Research Laboratory, Kirtland Air Force Base, NM 87117, USA}
\affil[e]{Optical Science \& Engineering Program, University of New Mexico, Albuquerque, NM 87131, USA}

\cftpagenumbersoff{figure}
\cftpagenumbersoff{table} 
\begin{document} 
\maketitle

\begin{abstract}
\Ac{GRIN} fibers are used to improve the design of many fiber optic devices.
However, the properties of the \ac{GRIN} fiber must be determined to
optimally engineer a device which incorporates \ac{GRIN} fiber components.
The index of refraction of most \ac{GRIN} fibers varies
quadratically in the radial direction, where the quadratic coefficient is characterized by the
gradient index constant $g$.
We measured $g$ for Thorlabs GIF50C \ac{GRIN} fiber at both
$780 \nm$ and $1550 \nm$ using equipment which is commonly available in an optics laboratory. This measurement was achieve by profiling the beam
exiting various lengths of \ac{GRIN} fiber.  A custom-built beam profiler
was used, which enabled the beam position to be referenced with respect to
the facet of the \ac{GRIN} fiber.  We report a gradient index constant of
$0.0057\um^{-1} \pm 0.0001\um^{-1}$ at $780\nm$ and $0.0055\um^{-1} \pm 0.0001\um^{-1}$ at $1550\nm$.  These results
are in close agreement with previously reported gradient index constant
measurements made for different wavelengths.
\end{abstract}


{\noindent \footnotesize\textbf{*}Spencer Olson,  \linkable{qst@afrl.af.mil} }

\acresetall
\section{Introduction}

\Ac{GRIN} optical fibers have a wide range of application including
space-division multiplexing~\cite{Richardson2013,Essiambre2013}, soliton
generation~\cite{Renninger2013,Ahsan2018,Sun2022}, and spatial-beam
cleanup~\cite{Liu2016,Krupa2017,Krupa2020}.  \Ac{GRIN} fiber may also be used to
form lenses which have been used to improve coupling into photonic integrated
circuits~\cite{Melkonyan2017}, fiber Fabry-Perot cavities~\cite{gulati_Fiber-Cavities_2017},
free-space-based fiber-optic components~\cite{vanBuren2003, Gomez-Reino2008},
as well as saturable absorbers used to create fiber-based lasers~\cite{Monroy2021}.

Understanding the refractive index profile, characterized by the gradient index constant $g$, is beneficial for many GRIN fiber applications.  In this
paper we present measurements of $g$ for
Thorlabs GIF50C \ac{GRIN} fiber at $780\nm$ and $1550\nm$.
Measurement of $g$ was achieved by profiling the Gaussian beam exiting
various length \ac{GRIN} lenses using a custom-built beam profiler.

\section{Theory}

For a typical \ac{GRIN} fiber, the refractive index as a function of radius
is modeled by~\cite{Wang:2011}
\begin{equation}
    n(r) = n_0\left(1 - \frac{g^2}{2}r^2\right)
\end{equation}
where $r$ is the radial distance from the center of the fiber, $n_0$ is the
index of refraction at the fiber center, and $g$ is the gradient index constant.
The period (or pitch) of the
\ac{GRIN} fiber is related to the gradient index constant by $P = 2\pi/g$.
A \ac{GRIN} fiber of gradient index constant $g$ and length $l$ is
characterized by the ABCD matrix~\cite{Wang:2011}
\begin{equation}
    M_g = \begin{pmatrix}
            \cos(g \, l)       & \frac{1}{g} \sin(g \, l)\\
            -g \, \sin(g \, l) & \cos(g \, l)
            \end{pmatrix}
\end{equation}

In the measurements presented here, we characterize light exiting a \ac{GRIN} lens
formed by splicing a small section of \ac{GRIN} fiber directly to a
\ac{SM} fiber.  For these \ac{GRIN} lenses, the Gaussian beam exiting the
\ac{GRIN}, $q_f$, is related to the Gaussian beam guided through
the \ac{SM} fiber, $q_0$, by the equation
\begin{equation}
    \label{eq:grin_len_model}
    q_f = M_r \, M_g \, q_0
\end{equation}
where $M_r$ is the ABCD matrix describing the Snell Law refraction which
occurs when light exits the \ac{GRIN} fiber.  We assume the \ac{GRIN} and
\ac{SM} fibers have minimal index mismatch, such that the refraction at the
interface of the two fibers may be ignored.  Using ABCD matrix analysis, the
gradient index constant $g$ may be calculated when $q_0$, $q_f$ and $l$ are
known.

\section{Gradient Index Constant Measurement}\label{app:grin-measurement}

The gradient index constant was determined through characterization of
beams exiting \ac{GRIN} lenses formed with known lengths of \ac{GRIN}
fiber.  The \ac{GRIN} lenses were fabricated in-house by fusion splicing
\ac{SM} fibers directly to \ac{GRIN} fibers.  The \ac{SM} fibers used
were Thorlabs PM780-HP for $780\nm$ and Thorlabs HB1550Z for $1550\nm$.
The \ac{GRIN} fiber was Thorlabs GIF50C.

Precise control over the initial length of the \ac{GRIN} fiber was
achieved through a modification to a fiber cleaver
(Fujikura CT-101), wherein the fiber holder was attached to a micrometer
translation stage.  This modification enabled control of the
relative distance between fiber cleaves.  Using this
technique, we achieved a $\pm10\um$ error between the \ac{GRIN}
length measured via a microscope and the micrometer reading.

With a \ac{GRIN} lens fabricated as described above, light was coupled into the
lens through the \ac{SM} fiber, and the light exiting the \ac{GRIN}
lens was characterized using a custom-built beam profiler.  The
profiler was built to enable the position of beam measurements to
be referenced to the location of the front facet of the fiber.  The beam profiler
consisted of an imaging system directed normal to the fiber facet.
Imaging was provided
by a camera along with imaging optics which produce $1.14\um$ pixel
size at $780\nm$ and $4.06\um$ pixel size at $1550\nm$.
Translation of the fiber along the imaging axis was achieved using
a closed-loop single-axis translation stage.  The imaging system
did not move during the measurements.

\begin{figure}[h]
    \centering
    \includesvg{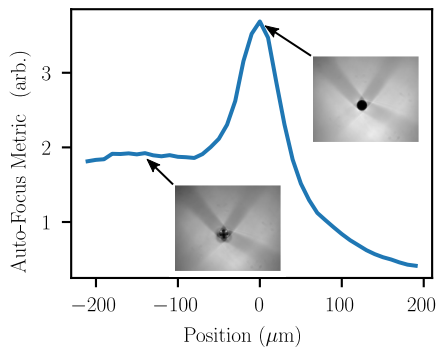}
    \caption{The auto-focusing metric versus fiber position.
             The pixel-brightness gradient in the image along the outside edge of the fiber tip was used as a
             metric to determine when the fiber was in the plane of focus. Inset
             images show the corresponding images of the fiber tip.}
    \label{fig:auto_focus}
\end{figure}

Image focusing was used to determine the location of the
fiber tip.  When the tip of the fiber was in-focus, a sharp
contrast was seen at the edges of the fiber.  The magnitude of
the two dimensional contrast gradient of the image measured along this
edge was used as a metric to determine when the fiber was in-focus.
A coarse auto-focusing algorithm determined the location of
the fiber tip to within $\pm25\um$.
This was followed by a fine scan centered about the coarse
fiber location estimate.  Images were captured at $10\um$ increments, and
the magnitude of the gradient was calculated for each image. To
minimize backlash, the fiber
always moved away from the camera during this auto-focusing
scan.  The location of the fiber was determined by fitting the
three largest gradient values to a quadratic, then calculating
the location of the maximum gradient according to this fit.  An illustration
of typical fine auto-focusing data is shown in Fig.~\ref{fig:auto_focus}.
Using this metric, the location of the fiber was determined
to within $\pm5\um$.

Once the location of the fiber was determined, the fiber was
moved so that the image plane was $0.5\mm$ in front of
the fiber tip.  Ten images of the light exiting the fiber were
averaged and fit to a two-dimensional Gaussian profile.
The distance between the fiber and the camera was increased in
$0.1\mm$ increments.  At each position, the beam was imaged and
a fit performed.  This process was repeated until the fiber tip has moved a total
of $1.9\mm$ from the image plane.

\begin{figure}[h]
    \centering
    \includesvg{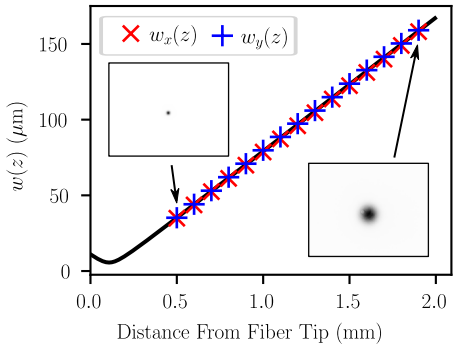}
    \caption{Beam size versus distance from fiber tip for $365.5 \um$
             length \ac{GRIN} and $1550\nm$ wavelength along with
             least-squares fit using Eq.~\ref{eq:rayleigh}.  Inset
             images show the corresponding beam image.}
    \label{fig:waist_vs_position}
\end{figure}

The horizontal and vertical beam widths were modeled to vary as a function
of distance according to
\begin{equation}
    \label{eq:rayleigh}
    w(z) = w_0\,\sqrt{1+\left( \frac{z-z_0}{z_R} \right)^2}
\end{equation}
where $w_0$ is the beam waist size and $z_0$ is the location of the
beam waist.  $z_R$ is the Rayleigh length, which
is given by $z_R= \pi w_0^2 \, n / \lambda$ where $\lambda$ is the
wavelength, and $n$ is the index of refraction of the air.
The beam width measurements along the vertical and horizontal axes
were fit to Eq.~\ref{eq:rayleigh}, where
$w_0$ and $z_0$ were the only free variables used during the fit.
The entire beam profiling measurement described above was repeated
for a total of five measurements.  An example of the beam width
data is shown in Fig.~\ref{fig:waist_vs_position}.

After measuring the Gaussian beam properties, the \ac{GRIN} lens
was polished using a standard fiber polisher, removing roughly
$50\um$ of material.  The entire
process for locating the fiber tip and measuring the beam waist
location and diameter was repeated.  The beam was profiled for a
total of fifty-one lengths of \ac{GRIN} for $780\nm$ light, and
forty-two lengths of \ac{GRIN} for $1550\nm$.

\begin{figure}[h]
    \centering
    \includesvg{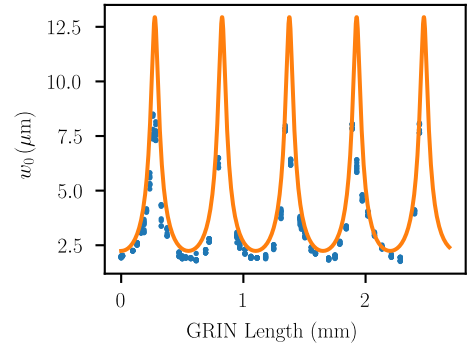}
    \caption{GRIN waist size data (dots) along with least-squares fit of the ABCD matrix model for 780\nm (line).}
    \label{fig:GRIN-w0_vs_length-780}
\end{figure}

\begin{figure}[h]
    \centering
    \includesvg{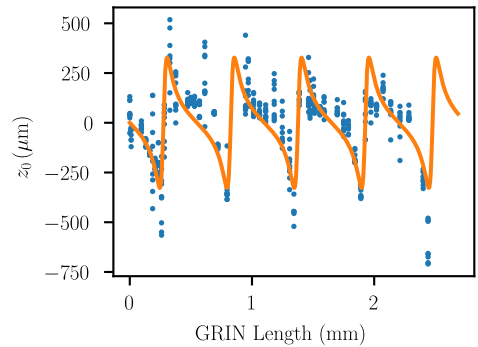}
    \caption{GRIN waist location data (dots) along with least-squares fit of the ABCD matrix model for 780\nm (line).}
    \label{fig:GRIN-z0_vs_length-780}
\end{figure}

\begin{figure}[h]
    \centering
    \includesvg{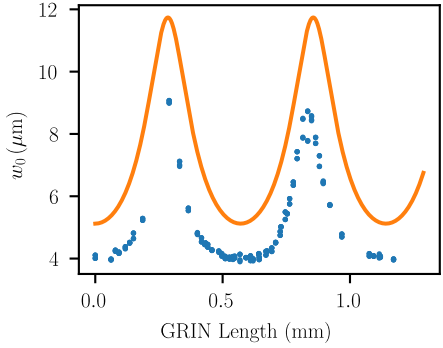}
    \caption{GRIN waist size data (dots) along with least-squares fit of the ABCD matrix model for 1550\nm (line).}
    \label{fig:GRIN-w0_vs_length-1550}
\end{figure}

\begin{figure}[h]
    \centering
    \includesvg{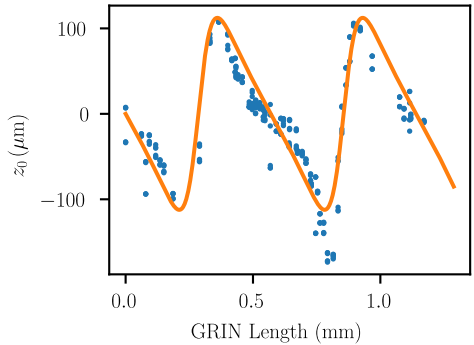}
    \caption{GRIN waist location data (dots) along with least-squares fit of the ABCD matrix model for 1550\nm (line).}
    \label{fig:GRIN-z0_vs_length-1550}
\end{figure}

Beam waist size and location data were fit to the quadratic index
variation model described in Eq.~\ref{eq:grin_len_model}.  The
beam waist inside the \ac{SM} fiber $w_0$ and the gradient
index constant $g$ were the only free-variables of the fit.  The
data along with the resulting fits are shown in
Figs.~\ref{fig:GRIN-w0_vs_length-780}, \ref{fig:GRIN-z0_vs_length-780},
\ref{fig:GRIN-w0_vs_length-1550}, and~\ref{fig:GRIN-z0_vs_length-1550}.
Through this analysis, the gradient index constant was found to be
$0.0057\um^{-1} \pm 0.00010\um^{-1}$ at $780\nm$ and $0.0055\um^{-1} \pm 0.0001\um^{-1}$ at $1550\nm$.
Although prior papers did not provide the model number of the \ac{GRIN}
fiber used and the work was for different wavelengths than presented here,
our calculated values for the gradient index constant closely match
the previously reported measurement of $0.00587\um^{-1}$
at 1310\nm~\cite{Wang:2016}, which was calculated via a refractive index profile tester, as
well as the value of $0.0055\um^{-1}$ at 1300\nm~\cite{Wang:2011}.

\pagebreak

Discrepancy between the model and the data seen in
Fig.~\ref{fig:GRIN-w0_vs_length-1550} can not be reduced through
variation of the free variables used during the least-squares fit,
as changes to the gradient index constant primarily manifest as
changes in the periodicity.  This suggest that the discrepancy between
the model and our data likely arises from systematic
measurement error or a model which does not adequately
describe our \ac{GRIN} lens.  The introduction of additional fit
parameters, such as a variable accounting for a systematic
error in the waist size measurements,
produces results which more closely match the data.
However, the covariance matrix calculated from these least-squares
fits shows that the gradient index constant is not strongly correlated
to other fit parameters.  Indeed, no change in the reported value
of $g$ was seen after increasing the number of free parameters.
Taken together, these results suggest that discrepancy between
our data and the model contribute little to the error in the
calculated gradient index constant.

\section{Conclusion}

We measured the gradient index constant of Thorlabs
GIF50C \ac{GRIN} fiber at both $780 \nm$ and $1550 \nm$ by profiling the Gaussian
beam exiting \ac{GRIN} lenses made with various lengths of \ac{GRIN} fiber.
This measurement was performed using equipment which is commonly available in an optics laboratory.
We report a gradient index constant of $0.0057\um^{-1} \pm 0.00010\um^{-1}$ at $780\nm$ and
$0.0055\um^{-1} \pm 0.0001\um^{-1}$ at $1550\nm$.  These values closely agree with previously reported
measurements, though for different wavelengths.  The determination of the gradient index constant may be used to
improve the design fiber optic devices through the incorporation of \ac{GRIN} components.

\section*{Disclaimer}
The views expressed are those of the authors and do not necessarily 
reflect the official policy or position of the Department of the Air 
Force, the Department of the Defense, or the U.S. Government. The
authors declare no conflicts of interest.

\subsection*{Disclosures}
The authors declare no potential conflicts of interest.

\subsection* {Data} 
Data underlying the results presented in this paper are not publicly
available at this time but may be obtained from the authors upon
reasonable request. 

\subsection* {Acknowledgments}
This work was funded by the Air Force Office of Scientific Research 
under lab task 22RVCOR017.


\bibliography{grin_characterization}
\bibliographystyle{spiejour}   

\end{document}